\documentstyle[proceedings,psfig]{crckapb10}
\begin{opening}
\title{OBSERVED REDSHIFT DISTRIBUTIONS AND\protect\\
       COSMOLOGICAL EVOLUTION OF RADIO SOURCES}
\author{J. MACHALSKI, M. JAMROZY}
\institute{Astronomical Observatory, Jagellonian University\\
           ul. Orla 171, PL-30244 Cracow, Poland}
\end{opening}

\runningtitle{REDSHIFT DISTRIBUTIONS}

\begin{document}

\begin{abstract}
It is shown that the new observed redshift distributions of various flux-limited 
samples of radio sources in general are consistent with the predictions of two 
basic evolutionary models published by Condon (1984) and Dunlop \& Peacock 
(1990), i.e. none of them can be rejected at the confidence level of about 95 
per cent. However, the models allowing a free-form evolution and suggesting both 
density and luminosity evolution are more consistent with the observational data 
\underline{at lower redshifts}, while the 'pure luminosity evolution' model fits 
better the data \underline{at higher redshifts}. This leads to a suspicion that 
the {\it same evolution} governing {\it all} radio sources, suggested by Condon 
(1984), might not be the case.
\end{abstract}
 

A lot of research was carried out during the last two decades to learn how the 
population of radio galaxies and quasars evolves at high redshift. One of the 
conclusions was that there is a 'redshift cutoff' in the distribution of 
{\it all} high-luminosity radio sources, in the sense of a decrease of their 
comoving space density over the redshift range 2--4. However, this is not clear 
whether the 'cutoff' and the decrease are the same for all evolving populations.

Two basic evolutionary models published by Condon (1984) and Dunlop \& Peacock 
(1990) [hereafter {\bf C84} and {\bf DP90} models] predict the redshift cutoff 
in different ways. The {\bf C84} model allows all sources to evolve in the same 
way, while the {\bf DP90} models (originally marked RLF1,2,3,4,5; Dunlop \& 
Peacock analyzed the data available at the time using different sets of starting 
parameters of the local {\it Radio Luminosity Function}) allow flat- and 
steep-spectrum sources to evolve independently. The {\bf DP90}, 'pure luminosity 
evolution' (PLE) model assumes a high-luminosity evolving population of 
elliptical galaxies, and a low-luminosity non-evolving population of 
spiral/irregular galaxies. The models of Condon and Dunlop \& Peacock differ 
also in accounting for flat- and steep-spectrum sources.


We made an effort to select a number of the flux-limited samples of radio 
sources for which redshift data are fairly complete or can be assumed to be 
representative. The old and new redshift samples are summarized in Table 1. 
The observed redshift distributions of these samples are displayed in Figures 
1 and 2.

\begin{table}[htb]
\begin{center}
\caption{The samples}
\begin{tabular}{lllllrrll}
\hline
Sample & Freq. & $F_{lim}$ & $\alpha$ & A & No. & No. & med. & Ref.\\
       & (GHz) & (Jy)      &       & (sr) & all & with z & z\\
\hline
PR+CJ1 & 5   & $\ge 0.7$  & $< 0.5$ & 1.84 & 108 & 97 & 1.08 & 5,6\\
       &     &            & $\ge 0.5$ &      &  92 & 90 & 0.40 & 5,6\\
HMIH   & 5   & 0.175--0.7 & $< 0.5$ & (0.4) & 114 & 68 & 1.45 & 2\\
WP     & 2.7 & $\ge 2.0$  & $< 0.5$ & 9.81 &  68 & 68 & 0.85 & 7\\
       &     &            & $\ge 0.5$ &      & 165 & 163 & 0.22 & 7\\
PKS    & 2.7 & $\ge 0.145$ & $< 0.5$ & 0.075 & 24 & 22 & 1.33 & 1\\
       &     &            & $\ge 0.5$ &      &  87 & 80 & 0.69 & 1 \\
GB/GB2 & 1.4 & $\ge 0.55$ & $< 0.75$ & 0.44 & 74 & 64 & 0.80 & 4\\
       &     &            & $\ge 0.75$ &     & 158 & 141 & 0.57 & 4\\
       & 1.4 & 0.2--0.55    & all     & 0.09 & 135 &  85 & 0.79 & 4\\
B2/B3  & 0.408 & $\ge 1.74$ & all     & 0.62 & 239 & 201 & 0.50 & 8\\
LRL    & 0.178 & $\ge 10.9$ & all     & 4.05  & 181 & 179 & 0.48 & 3\\
\hline
\end{tabular}
\end{center}
\begin{minipage}{10cm}
\hspace{1cm}{\bf References to Table 1:}\\
\small{
\hspace*{1cm}(1) Downes, Peacock, Savage \& Carrie, 1986, MNRAS,218,31\\
\hspace*{1cm}(2) Hook, McMahon, Irwin \& Hazard, 1996, MNRAS,282,1274\\
\hspace*{1cm}(3) Laing, Riley \& Longair, 1983, MNRAS,204,151\\
\hspace*{1cm}(4) Machalski, 1997, A\&AS (submitt.)\\
\hspace*{1cm}(5) Pearson \& Readhead, 1988, ApJ,328,114\\
\hspace*{1cm}(6) Polatidis et al., 1995, ApJS,98,1\\
\hspace*{1cm}(7) Wall \& Peacock, 1985, MNRAS,216,173\\
\hspace*{1cm}(8) this paper}
\end{minipage}
\end{table}

\begin{figure}
\centerline{\psfig{figure=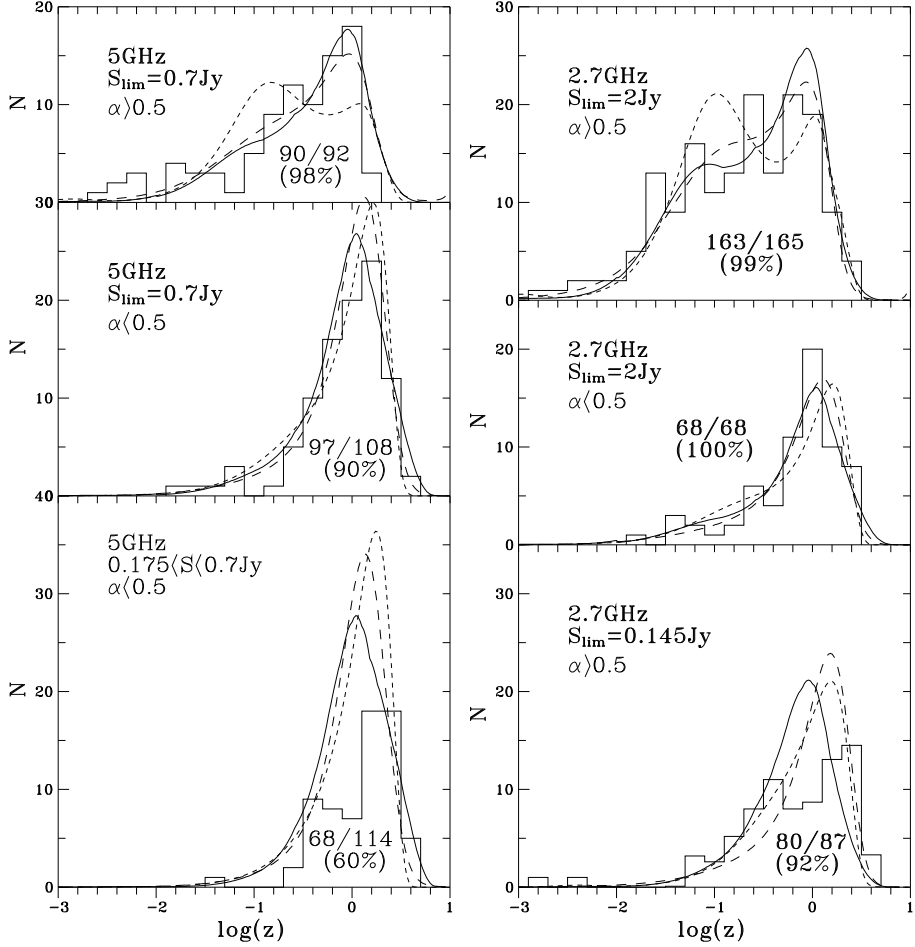,width=12cm}}
\caption{\label{fig1}
{\bf Left panel:} The n(z) distributions at 5 GHz. The solid curves show the 
distributions predicted from the C84 model; the long-dash and short-dash 
curves -- those from the RLF3 and PLE models, respectively. {\bf Right panel:} 
the same as in left panel but at 2.7 GHz}
\end{figure}


The n(z) distributions, predicted from the C84 as well as RLF3 and PLE models, 
are shown as continuous curves overlayed on the observed redshift distribution 
of particular samples. The latter two models are chosen because of their very 
discrepant predictions. Though the division of sources into {\it 'flat-spectrum'}
and {\it 'steep-spectrum'} populations, in the face of {\it unified models}, is 
strongly unjustified -- we retain this division because the models were 
originally designed this way and can be used to predict n(z) distributions for 
sources with different spectra. At the frequencies of 5 and 2.7 GHz the 
distributions are separated by $\alpha =0.5$, while at 1.4 GHz -- $\alpha =0.75$ 
was chosen.

\begin{figure}
\centerline{\psfig{figure=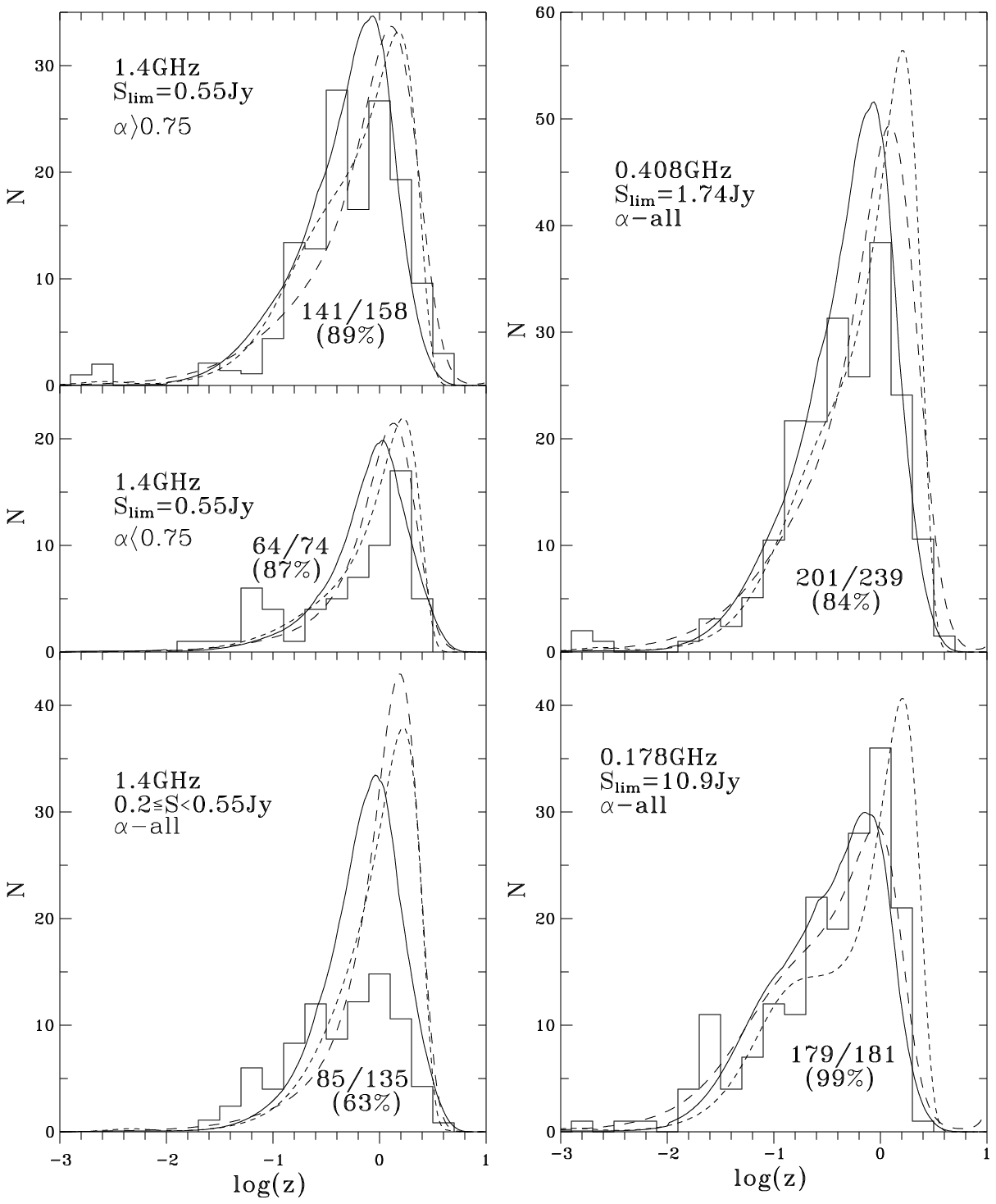,width=12cm}}
\caption{\label{fig2}
{\bf Left panel:} The same as in Fig.1 but at 1.4 GHz. {\bf Right panel:} 
the same as in Fig.1 but at 408 and 178 MHz}
\end{figure}


The goodness-of-fit (chi-squared, Kolmogorov-Smirnov) tests applied to the 
observed and predicted distributions have shown that
\noindent
(1) None of the models can be rejected at the confidence level of about 
95 per cent.
\noindent
(2) Considering all available n(z) distributions, the C84 model fits the 
observations better than do this the DP90 RLF1--5 and PLE models.
\noindent
(3) The PLE model fits better the redshift data of the samples with lower 
flux limits, i.e. comprising sources at higher redshifts. The C84 model fits 
better the data of the strong samples (with sources at lower redshifts) where 
the PLE model cannot be accepted (cf. n(z) distributions of steep-spectrum 
sources in the 5 GHz 0.7Jy and 2.7 GHz 2Jy samples).

Two conclusions can be drawn from the above analysis:

1. The evolutionary models allowing both density and luminosity evolution (C84, 
RLF3) are more consistent with the observational data at lower redshifts, while 
the PLE model fits better the data at higher redshifts. This is not consistent 
with the scheme in which all radio sources evolve in the same way (e.g. Condon 
1984). 

2. All redshift-complete samples, available up to date, probe almost the same 
radio luminosity range (the data in Table 1 show overall increase of the median 
redshift with a decrease of the flux limit at each observing frequency). 
Therefore, these data are still not sufficient to determine whether the 
high-redshift cutoff of the form manifested by luminous sources holds for less 
luminous sources, and/or is different for separable populations, e.g. 
different FR-type  radio galaxies, quasars, CSS sources, GPS sources, etc. 
Future, more complete, redshift data at the flux levels of about 1 -- 5 
mJy should improve our knowledge about space distribution of radio sources and 
its evolution.

\end{document}